\documentclass{jetpl}
\twocolumn
\newcommand{\be}{\begin{eqnarray}}
\newcommand{\ee}{\end{eqnarray}}
\newcommand{\ba}{\begin{array}}
\newcommand{\ea}{\end{array}}

\lat

\title{On the Pion Distribution Amplitude Shape}

\rtitle{On the Pion Distribution Amplitude Shape}

\sodtitle{On the Pion Distribution Amplitude Shape}

\author{M.V.~Polyakov$^{1,2}$}

\rauthor{M.V.~Polyakov}

\sodauthor{ M.V.~Polyakov}

\address{$^1$ Institut f\"ur Theoretische Physik II,
Ruhr--Universit\"at Bochum, D--44780 Bochum, Germany\\
$^2$Petersburg  Nuclear Physics Institute, Gatchina, 188300, St.
Petersburg, Russia}

\abstract{We argue that the recent BaBar data on $\gamma \to \pi$ e.m. transition form factor at large photon
virtuality supports the idea that pion distribution amplitude (DA) is close to unity with $\phi_\pi^\prime(0)/6\gg 1$
at a normalization point of $\mu=0.6\div 0.8$~GeV.
Such pion DA can be obtained in the effective chiral quark model. The possible flat shape of the pion DA implies that
the standard expansion of the DA in Gegenbauer polynomials can be divergent.

On basis of chiral models we predict that the two-pion DA should exhibit
anomalous endpoint behaviour for pions in the S-wave and that such feature is
absent for higher partial waves. The latter implies that the $\rho,
f_2$, etc. meson DAs have no anomalous endpoint behaviour.
Possible implications of such pion DA for other hard exclusive
processes are shortly discussed.}

\begin{document}

\maketitle
%\PACS{13.60.Le\and14.20.Gk}

The reaction $\gamma^*(q)+\gamma(q')\to \pi^0$ at large  virtuality of
the photon ($-q^2=Q^2\gg \Lambda_{\rm QCD}^2$) is the basic {\it hard exclusive}
process which allows rigorous QCD description \cite{pionery}.  QCD allows to
make remarkably nice
prediction for the $Q^2\to\infty$ limit of the $\gamma\pi$ transition form factor:

\be \label{asympt} \lim_{Q^2\to\infty}Q^2
F_{\gamma\pi}\left(Q^2\right)=2f_\pi\, . \ee
Here $f_\pi \approx
0.0923$~GeV is the pion decay constant.
The approach to the asymptotic (\ref{asympt}) can be written to the leading order in $\alpha_s(Q^2)$
as follows\footnote{Note, that in Eq.~(\ref{predasy}) we
modified perturbative quark propagator $1/(z\ Q^2)$ by $1/(z\ Q^2+m^2)$, where $m$ stays for possible
non-perturbative contributions to the quark propagator. We stress that
we do not derive such a modification, but use it just to mimic non-perturbative
contributions. Such simplification is enough for rather qualitative discussion here.
Derivation of the modification of the quark propagator due to e.g. instanton
non-perturbative contribution will be given elsewhere.} \cite{pionery}:

\be
\label{predasy}
Q^2
F_{\gamma\pi}\left(Q^2\right)=\frac{2f_\pi}{3}\ \int_0^1 dz\ \frac{\phi_\pi(z,Q)}{z +m^2/Q^2}+O(\alpha_s(Q^2)).
\ee
Here $\phi_\pi(z,\mu)$ is the pion distribution amplitude (DA) at the normalization scale $\mu$. It is defined as the
following matrix element:

\be
\label{pionDA}
&&\langle 0 | \bar d (n) \gamma_\mu n^\mu \gamma_5 [n, -n] u (-n)
| \pi^+ (P ) \rangle\\
\nonumber
&=& i \sqrt{2} f_\pi (n\cdot P)
\int_0^1 dz \; e^{i (2 z - 1) P\cdot n} \phi_\pi (z)
\label{phi_pion}
\ee
Here $n^\mu$ is a light--like 4--vector ($n^2 = 0$), and
\be
[n, -n] &\equiv& \mbox{P}\, \exp \left[
\int_{-1}^1 dt\; n^\mu A_\mu (t n)
\right]
\label{P_exp}
\ee
denotes the path--ordered exponential of the gauge field, required by
gauge invariance; the path is defined to be along the light--like
direction $n$.

The pion DA can be represented as the series in the eigenfunctions of the leading order
evolution equation -- Gegenbauer polynomials \cite{pionery}:

\be
\label{gexp}
\phi_\pi(z,\mu)=6 z(1-z)\left(1+a_2(\mu) C_2^{\frac 32}(2 z-1)+\ldots \right)\, .
\ee
Usually it is tacitly assumed that the series (\ref{gexp}) is convergent, that is why in analyses
of experimental data (see e.g. \cite{Kroll:1996jx,Schmedding:1999ap,Bakulev:2002uc})
only a finite number of terms in this series is considered. Actually, the assumption
about the convergence of the series (\ref{gexp})
does not follow from {\it any} principle. Moreover, there are counterexamples for such convergency. First
counterexample results from the effective chiral quark model calculations of the photon DA in  \cite{ppgw}. It was shown
that the photon DA is not zero at the endpoints $z=0,1$, which  explicitly demonstrated that the series (\ref{gexp})
is divergent. The result for the pion DA in the same model \cite{ppgw,pp} is $\phi_\pi(z,\mu_0)=1$
at the normalization point $\mu_0=1/\rho\sim 0.6$~GeV determined by the average size of the instanton.
The normalization scale in the chiral effective quark model is inherited from the theory of instanton vacuum \cite{diapet,dpw}
from which the effective quark model has been derived. However, it was noted in \cite{off} that the effective quark model
should be extended to the higher orders in instanton packing fraction $\rho^2/R^2\sim 1/10$ when one  considers pion DA
at $z\sim \rho^2/R^2$. In Refs.~\cite{ppgw,pp,Anikin:2000bn,Praszalowicz:2001wy,Nam:2006sx} it was suggested {\it ad hoc} modification of the model beyond the leading order
in the instanton packing fraction which led to the pion DA only slightly wider than the asymptotic one (however rather strong sensitivity to additional {\it ad hoc}
parameters was demonstrated). The proposed in Refs.~\cite{ppgw,pp,Anikin:2000bn,Praszalowicz:2001wy,Nam:2006sx} extensions of the effective chiral quark model were based
essentially on modelling the momentum dependence of the dynamical quark mass $M(p)$ by the rational function of the momentum. A drawback of such extensions is that the
endpoint nullification
of the pion DA rely on contribution of the artificial (non-physical) poles in the used Ansatz for $M(p)$. Position of that poles
is far from the Euclidean
domain where one can trust the result of the instanton liquid model.
Another problem of the {\it ad hoc} modification of the model beyond the leading order
in the instanton packing fraction used in Refs.~\cite{ppgw,pp,Anikin:2000bn,Praszalowicz:2001wy,Nam:2006sx} is that the axial current
is not conserved to the order $\rho^2/R^2$. Possible solution of the problem with the axial current conservation was suggested in Ref.~\cite{Bzdak:2003qe}.
The calculation of the pion DA with improved axial current in Ref.~\cite{Bzdak:2003qe} gave the function which is non-zero at the endpoints.

We can summarize that the theory of the instanton vacuum in the leading order of the instanton packing fraction predicts $\phi_\pi(z,\sim 1/\rho)=1$, however
in the vicinity of the endpoint of order $\rho^2/R^2$ theory should be modified. Precise form of the modification is not strictly derived, that is very interesting
problem to study.
Given such state of art, we can only state that
 the pion DA in the instanton theory of QCD vacuum is expected to be rather flat -- meaning that it is close to unity with $\phi_\pi^\prime(0)/6\gg 1$.

  Calculation of the pion DA in the Nambu--Jona-Lasinio model \cite{RuizArriola:2002bp} gave the same result as in the leading order
effective chiral quark model -- $\phi_\pi(z,\mu_0)=1$, however it was attributed to a very low normalization point of $\mu_0=0.313$~GeV.
The same result is obtained in the large--$N_c$ Regge model \cite{RA}.

We see that the wide class of chiral models predict the pion DA which is flat and even non-zero at the end points. We note that the possibility
of the pion DA $\phi_\pi(z)=1$ was considered almost three decades ago in Ref.~\cite{Efremov:1980mb}. Recent studies of the hadronic wave function in AdS/QCD
\cite{Brodsky:2006uqa} also suggests the wide pion DA $\phi_\pi(z)\sim \sqrt{z(1-z)}$ with anomalous behaviour at the endpoints,
supporting the idea that the series (\ref{gexp}) is divergent.

In these notes we consider an extreme possibility that the pion DA is $\phi_\pi(z,\mu_0)=1$ at a normalization point of $\mu_0\approx 0.6-0.8$~GeV.
The same shape as in the instanton liquid
 model in the leading order in the instanton packing fraction $\rho^2/R^2$ \cite{ppgw,pp}. Such assumption about pion DA
would imply that the scaled form factor
$Q^2 F_{\gamma\pi}(Q)$ overcomes the asymptotic value given by Eq.~(\ref{asympt}) and then very slowly approaches it from above\footnote{ Interestingly, such enhancement
for the similar to $\gamma^*\gamma\to \pi^0$ process--
DVCS amplitude-- was discussed in Ref.~\cite{off} where the generalized parton distributions were computed in the effective chiral quark model and
the anomalous endpoint behaviour was found.}.

 Recently the BaBar collaboration reported \cite{babar} results for the scaled form factor $Q^2 F_{\gamma\pi}(Q)$
for $4<Q^2<40$~GeV$^2$. Despite common expectations \cite{Kroll:1996jx,Schmedding:1999ap,Bakulev:2002uc} the scaled form factor crosses the asymptotic line
of $2 f_\pi$ around $Q^2=10$~GeV$^2$ and continues to grow slowly at higher $Q^2$. The BaBar data \cite{babar} are shown in Fig.~1.

\begin{figure}
\includegraphics[width =7.cm]{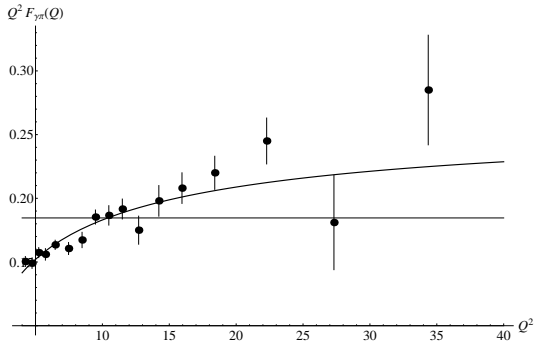}
\caption{Fig.~1. The scaled form factor $Q^2 F_{\gamma\pi}(Q)$ as a function of $Q^2$. Points with error bars are BaBar data \cite{babar}, thick solid line
our fit with pion DA (\ref{nparam}), horizontal line is the QCD asymptotic value (\ref{asympt}). } \label{fig:halla}
\end{figure}

Now we make a following simple exercise. We assume that the shape of pion DA at the normalization point $\mu_0=0.6 \div 0.8$~GeV has the following form:
\be
\label{nparam}
\phi_\pi(z,\mu_0)=N+(1-N) 6 z(1-z)\, ,
\ee
where $N$ is a free constant. We evolve the DA (\ref{nparam}) to the scale of $Q^2$ and then vary parameter $N$ in Eq.~(\ref{nparam}) and mass parameter $m$ in Eq.~(\ref{predasy})
to fit the BaBar data \cite{babar}. The obtained values of the parameters are:
\be
\label{values}
N=1.3\pm 0.2, \ \ \ m=0.65 \pm 0.05~ {\rm GeV}\,,
\ee
indicating that the BaBar  data favour the flat pion distribution amplitude.

Notably the resulting value of the mass parameter $m$ is close to
the inverse instanton size, which sets the scale for non-perturbative effects in quark propagator. The contribution of the pion DA (\ref{nparam}) with
the central values of parameters (\ref{values}) to the scaled form factor is shown by the  thick solid line in Fig.~1.

We note that our ``back of envelope" analysis is rather oversimplified as it is limited to the leading order in $\alpha_s(Q^2)$ and we used rather simple model for the
higher twists (\ref{predasy}). We believe that such model for the higher twist contributions, in the case of a flat pion DA, grasps the most essential
contributions related to the non-perturbative contributions to the quark propagator.

Qualitatively speaking, the BaBar data show that the rise  of the scaled
form factor with $Q^2$ for $Q^2>10$~GeV$^2$ is rather slow, indicating that the $Q^2$ dependence is governed by rather large mass parameter of order of several GeV. The way to obtain
such large mass scale from rather low non-perturbative mass scales of order of hundreds MeV is to enhance the latter due to the endpoint contribution of the flat
pion DA. Such ``transmutation" of mass scales is taken into account by our simple formula (\ref{predasy}) in which the low non-perturbative mass scale
$m\sim 0.65$~GeV is transformed into the large mass scale characteristic for the $Q^2$ dependence of the scaled form factor observed by the BaBar collaboration.

We are also limited ourselves to the leading order in
$\alpha_s(Q^2)$, for the flat pion DA the next-to-leading (NLO)
contributions \cite{nlo} can be large. Possible large NLO
corrections indicates that for the complete QCD analysis one needs
some kind of  resummation of the higher order contributions or
modifications of the NLO coefficient function.
 That is very interesting problem one can study in future.

The pion DA (\ref{nparam}) should not be taken literary, its form is just a handy way to parametrize a class of flat functions (close to unity and
with $\phi_\pi^\prime(0)/6\gg 1$). In principle one can try the functional form  $\phi_\pi(z)\sim [z(1-z)]^\alpha$ with $\alpha\ll 1$.
Let us point out that
the pion DA (\ref{nparam}) has the following coefficients in the Gegenbaur expansion (\ref{gexp}):

\be
a_2(2\ {\rm GeV})\approx 0.3, \ \ a_4(2\  {\rm GeV})\approx 0.15 .
\ee
These values are compatible with estimates by various non-perturbative methods, see survey of the results
in \cite{Bakulev:2002uc}. The main difference of our analysis with the commonly accepted that (see e.g. \cite{Kroll:1996jx,Schmedding:1999ap,Bakulev:2002uc})
is that we {\it do not assume} that the Gegenbauer series (\ref{gexp}) is convergent.

We were motivated to try the flat pion DA against the new BaBar data by the results of the instanton liquid model of QCD vacuum, which predicts
in the leading order of the instanton packing fraction $\phi_\pi(z)=1$ at
the normalization point of $\mu_0\sim 1/\rho\sim 0.6$~GeV.
Such unusual picture of pion DA has direct implications for other hadronic DA. In particular, the
two-pion DA \cite{diehl} (entering QCD description of hard $\gamma^*\gamma\to 2\pi$ processes) computed in the instanton liquid model of QCD vacuum has the
following form (in the leading order in the instanton packing fraction and at $m_{\pi\pi}=0$) \cite{pw}:

\be
\label{2pi}
&&\phi_{2\pi}(z,\zeta)=-(2z-1)+\left[2 z \theta(1<z<\zeta)\right.\\
\nonumber
&+&(2 z-1) \theta(\zeta<z<1-\zeta) +2 (z-1) \theta(1-\zeta<z<1)\left.\right].
\ee
Note that the first term in the above equation originates from {\it the contact} two pion couplings to quarks required by the
spontaneously broken chiral symmetry. The remarkable feature the first term is that
it is non-zero at the endpoints. The following terms in Eq.~(\ref{2pi}) are zero at the endpoints. The presence of the first term implies that
the scaled amplitude of the $\gamma^*\gamma\to 2\pi$ given by the formula similar to Eq.~(\ref{predasy}) should exhibit the $Q^2$ rise as for the
$Q^2 F_{\gamma\pi}(Q)$. However, there is an important difference -- the term with endpoint singularities in Eq.~(\ref{2pi}) is $\zeta$ independent.
That means that the raise of the amplitude with $Q^2$ is expected only for two pions in the S-wave. The two pion DA $\phi_{2\pi}(z,\zeta)$
in other partial waves is expected to be free from the endpoint singularities. Using the connection of two pion DA with the DAs of the resonances \cite{p99}
we predict that DAs of mesons with non-zero spin ($\rho, f_2$, etc.) are not anomalously flat as the pion DA.

Physics picture behind the flat pion DA can be traced back to Nambu--Goldstone nature of the pion. Due to the spontaneous breakdown
of the chiral symmetry in QCD the quark acquires sizable mass and in the hadron spectrum contains (almost) massless Nambu-Goldstone bosons (pions).
The broken chiral symmetry dictates that the $\pi q\bar q$ coupling is proportional to dynamical quark mass and is rather large ($g_{\pi q\bar q}\sim M/f_\pi$).
The instanton mechanism for chiral symmetry breaking \cite{diapet,dpw} predicts that this coupling is almost {\it point-like}--meaning that it is
rather sizeable for $k_\perp$ of quark up to momenta $\sim 1-2$~GeV. Presence of such ``point-like" component in the pion is the reason for the
flat pion DA. Possible existence of the ``point-like" component can have consequences for other hard processes, for instance, it can contribute
considerably to hard exclusive pion production off nucleon  at $Q^2$ of order several GeV$^2$. Such contribution is obviously sensitive to the chirally odd
generalized quark distributions in the nucleon -- transversity distributions. We note also the point-like coupling of Nambu--Goldstone
bosons to quarks appears also in V.N.~Gribov theory of quark confinement \cite{BH}.

Finally we note that the phenomena, similar to the anomalous endpoint behaviour of the pion DA
has been discussed in the chiral models for nucleon GPDs \cite{off} and nucleon DAs \cite{Petrov:2002jr}.

Due to big importance of the phenomenon observed by the BaBar collaboration, these results should be cross checked by other independent
experimental groups.

\section*{Acknowledgements}
I am grateful to N.~Kivel, D.~M\"uller and M.~Strikman for illuminating
discussions.
Especially I enjoyed discussions with A.~Radyushkin,
who independently and from different perspective
came to the conclusions that the BaBar data favour the flat pion DA \cite{tolya}. I am thankful to him for exchange by
ideas and his advises.

%%%%%%%%%%%%%%%%%%%%%%%%%%%%%%%%%%%%%%%%%%%%%%%%%%%%%%

\end{document}